\documentclass[reprint,aip,jap]{revtex4-1}
\usepackage{graphicx, dcolumn, bm, xcolor}
\usepackage{textcomp}
\usepackage{amsmath}
\usepackage{natbib}
\usepackage{ulem} 

\begin{document}
\title{Limits of emission quantum yield determination}

\author{Bart van Dam}
\affiliation{University of Amsterdam, Institute of Physics, Science Park 904, 1098 XH Amsterdam, The Netherlands}
\author{Benjamin Bruhn}
\affiliation{University of Amsterdam, Institute of Physics, Science Park 904, 1098 XH Amsterdam, The Netherlands}
\author{Gejza Dohnal}
\affiliation{Czech Technical University in Prague, Faculty of Mechanical Engineering, Karlovo n\'{a}m\v{e}st\'{i} 13, 121 35, Prague 2, Czech Republic  }
\author{Kate\v{r}ina Dohnalov\'{a}}
\affiliation{University of Amsterdam, Institute of Physics, Science Park 904, 1098 XH Amsterdam, The Netherlands}
\altaffiliation[Presently]{K. Newell}
\email{k.newell@uva.nl}
\date{\today}

\begin{abstract}
The development of new fluorescent molecules and dyes requires precise determination of their emission efficiency, which ultimately defines the potential of the developed materials. For this, the photoluminescence quantum yield (QY) is commonly used, given by the ratio of the number of emitted and absorbed photons, where the latter can be determined by subtraction of the transmitted signal by the sample and by a blank reference. In this work, we show that when the measurement uncertainty is larger than 10\% of the absorptance of the sample, the QY distribution function becomes skewed, which can result in underestimated QY values by more than 200\%. We demonstrate this effect in great detail by simulation of the QY methodology that implements an integrating sphere, which is widely used commercially and for research. Based on our simulations, we show that this effect arises from the non-linear propagation of the measurement uncertainties.  The observed effect applies to the measurement of any variable defined as $Z=X/Y$, with $Y=U-V$, where  $X$, $U$ and $V$ are random, normally distributed parameters. For this general case, we derive the analytical expression and quantify the range in which the effect can be avoided.
\end{abstract}

\maketitle

\section{Introduction}
The photoluminescence quantum yield (QY) is commonly applied to quantify the emission efficiency of fluorescent molecules and dyes in their development for lighting applications. The QY is given by the ratio of the number of emitted $N_{em}$ and absorbed $N_{abs}$ photons, where the absorption is commonly obtained by comparison of the total number of photons transmitted by the sample ($N_S$) and a blank reference ($N_{Ref}$): 
\begin{eqnarray}
QY=\frac{N_{em}}{N_{abs}}=\frac{N_{em}}{N_{Ref}-N_{S}} \label{eqQYbasic}
\end{eqnarray}
The QY is widely used also in research: In the past decade, research on quantum dot `phosphors' has been relying on the QY methodology to show various size- \cite{Mastronardi2012, Miller2012, Sun2015}, excitation- \cite{Timmerman2011, Saeed2014, Chung2017, Valenta2014} or concentration-dependent properties \cite{Timmerman2011, Greben2015}. Several guidelines exist for the QY measurements,\cite{Wurth2013, Valenta2014a} discussing e.g.\ the effects of re-absorption \cite{Ahn2007} and excitation geometry \cite{Faulkner2012, Wurth2015a}. However, none of them reflect on the critical effect of low sample absorption in the presence of measurement uncertainty - which can lead to dramatically biased results, as will be discussed here.

In this work we show that when the magnitude of the difference between the signals of the reference and sample measurement, used to evaluate the QY, reach levels comparable to that of experimental uncertainty (e.g.\ noise), the QY value can be heavily underestimated. This happens already under common experimental conditions, especially when absorption of the sample is relatively weak. For example, for noise levels of $\sim1\%$ the QY estimates for samples with an absorptance\footnote{Defined as the fraction of the incident light that is absorbed.} below $\sim10\%$ are affected. The effect is analyzed here in a great detail by simulating the QY methodology using an analytical model with careful consideration of the involved measurement uncertainties. We show that the uncertainty in the variables used to estimate the QY propagates in a non-linear way and leads, not only to a larger uncertainty in the QY value as is commonly assumed, but also to a considerable underestimation of the most-likely observed QY. This underestimation arises purely from the form in which the QY is defined, given by the general relation $X/Y$, and shows in case the denominator $Y=(U-V)$ is small. Therefore, our findings can be extended to any quantity determined by such type of a relation. 

\begin{figure*}[ht]
\centering
\includegraphics[width=0.7\textwidth]{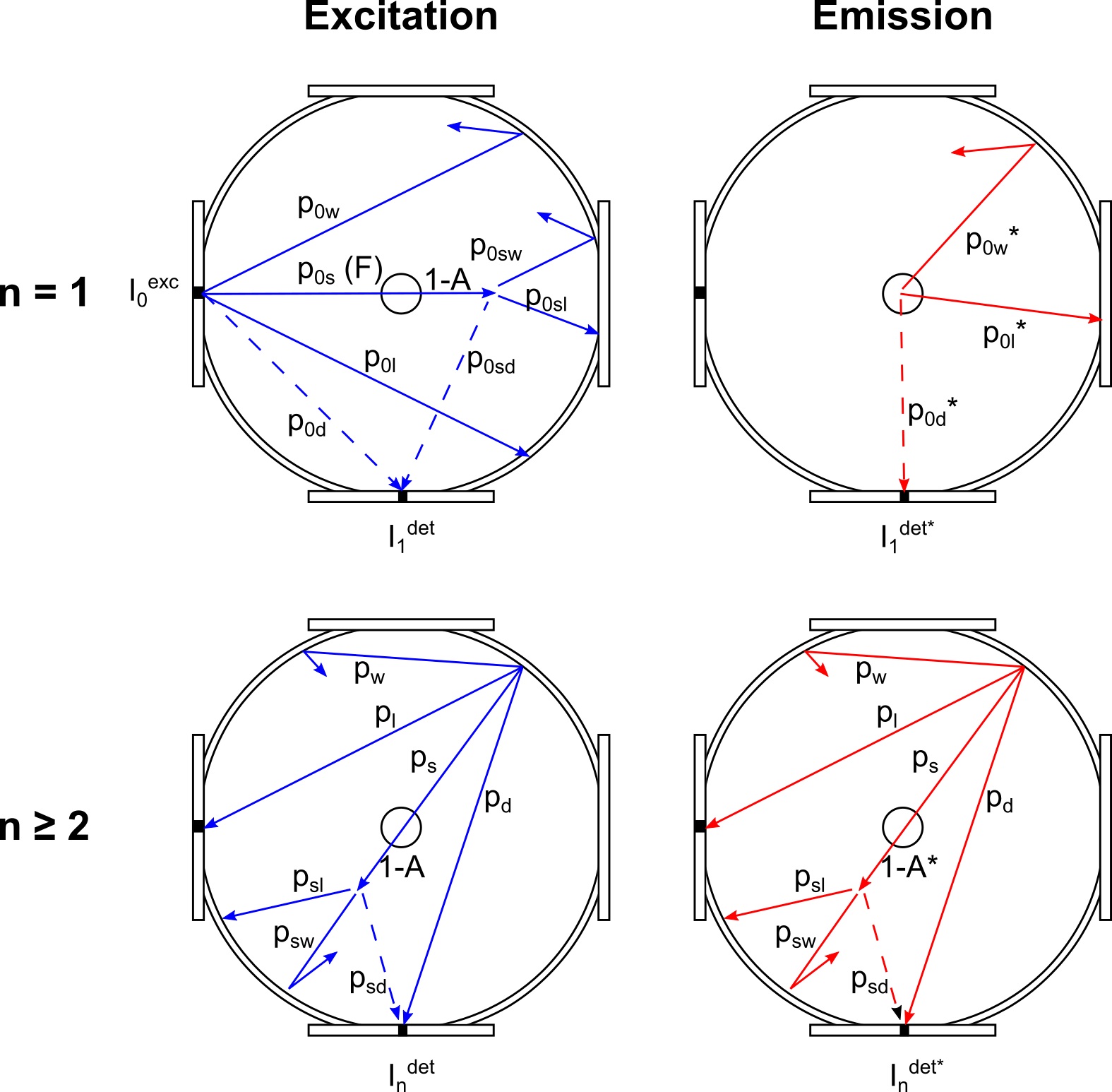}
\caption{(Color online) Schematics of a generalized integrating sphere (IS) setup used for our model. The IS cavity has a small entrance on the left side and an exit on the bottom side, where a detector is placed. The sample/reference is modelled as a spherical object suspended in the center. Lines represent the light paths between different objects inside the IS - wall (w), loss channel (l), detector (d) and sample (s). The parameters represent the probabilities of taking each specific light path. Light paths shown in dashed lines are prevented by the screen (so called 'baffles'). The light paths before the first reflection from the IS wall ($n=1$) are considered separately to account for the inhomogeneity of the light field in that instance. This differs for excitation light (entering from outside) and emission (originating from the center). Emission (red lines) is assumed separately from the excitation (blue lines) due to different spectral ranges, where the reflectivity of the sphere and the sample's absorption differ.}
\label{figure1}
\end{figure*}

\section{QY methodology}\label{QYmethodology}

We choose to demonstrate this effect on the QY method. To this end, we developed an analytical model to simulate the QY experiment. For this we select the optical QY methodology introduced by  \textit{Greenham} \cite{Greenham95} and \textit{de Mello et al.}\ \cite{DeMello1997}, and simplified by \textit{Mangolini et al.} \cite{Mangolini2006}. This technique implements an integrating sphere (IS), a reflectively scattering cavity, that allows the determination of the absolute number of emitted and absorbed photons, without the need for a QY calibration standard. The use of the IS has been standardized for LED and display industry, and QY devices based on the IS methodology are commercially available \cite{HamamatsuQY, HoribaQY}. Our model assumes a simplified and generalized IS geometry as shown in Figure~\ref{figure1}, implementing the experimental scheme described by \textit{Mangolini et al.} \cite{Mangolini2006}. In this method, a sample (e.g.\ a cuvette containing a solvent in which emitting nanoparticles are dispersed) is suspended inside the IS and excited from the entrance port. The excitation photons that are absorbed by the sample, are subsequently emitted at a different wavelength, with a probability given by the quantum efficiency $\eta$. After multiple reflections and scattering events within the IS, the excitation and emission photons are ultimately detected, lost or (re-) absorbed in the sample. The same measurement is repeated with a blank reference sample (e.g.\ the cuvette with solvent). From the difference in detected photon intensities at the excitation wavelengths ($I$) and emission wavelengths ($I^*$), the numbers of absorbed and emitted photons are evaluated:
\begin{eqnarray}
QY=\frac{N_{em}}{N_{abs}}=\frac{N_S^*-N_{Ref}^*}{N_{Ref}-N_S}=\frac{\int(I_{S}^{*}-I_{Ref}^{*})C^*\text{d}t}{\int(I_{Ref}-I_{S})C\text{d}t},
\label{eq1}
\end{eqnarray}
where subscripts $S$ and $Ref$ refer to the sample and reference measurements. The factors $C$ and $C^*$ correct for the spectral sensitivity of the detector and IS at the excitation and emission wavelength, respectively. For this an additional calibration measurement is performed using an empty IS and a calibration source with a known spectrum. \\

\section{Analytical model}\label{model}
We model the IS as a spherical cavity with two small openings: an entrance port from which excitation light enters and an exit port equipped with a detector. The sample or reference is suspended in the middle of the IS and is modeled as a spherical object with absorptance $A$,  reflectance $R$ and transmittance $T$, where $A+R+T=1$. The interior of the IS is covered by a coating that is highly reflective over a broad spectral range $R_w$ (usually $>97\%$) and acts as an ideal scatterer, i.e. the directionality of the light is lost after a single reflection from the walls. We define the probability $p$ that a photon impinges on an object inside the IS by the relative area of the object to the area of the IS interior. For example the probability of hitting the wall ($p_w$), loss channel ($p_l$), detector ($p_d$) or sample ($p_s$), with $1=p_l+p_d+p_s+p_w$. Since the ideally scattering walls ensure spatially distribution of the photons, we assumed that these probabilities do not depend on the exact location in the IS at which the photon scatters. However, to take into account that all the light starts from a single point (i.e. excitation from the entrance port and emission from the sample), we separate the first round of light reflection ($n=1$) from the consequent ones ($n\geq2$) (see Figure~\ref{figure1}). We do this by assigning modified probabilities $p_{0x}$ of hitting objects inside the IS, given by their visibility from the entrance. Again, $1=p_{0l}+p_{0d}+p_{0s}+p_{0w}$  and $1=p_{0w}^*+p_{0l}^*+p_{0d}^*$. $p_{0s}$ represents the fraction of the initial excitation light intensity, $I_0^{exc}$, that hits the sample directly, a parameter that is commonly varied in literature \cite{Faulkner2012,DeMello1997,Wurth2015a}. To separate this parameter from the other probabilities, we set $p_{0s}=F$, where $F=1$ for direct or $F=0$ for indirect excitation conditions. Furthermore, in accordance with the standard IS methodology, direct detection of the excitation and emission photons is prevented by an inserted baffle by setting $p_{0sd}=p_{sd}=p_{0d}=p_{0d}^*=0$ (dashed lines in Figure~\ref{figure1}). For the sake of completeness, we finally assume that the measurement is in a regime in which the QY is independent of the excitation photon flux.

From the light paths illustrated in Figure~\ref{figure1} and their probabilities $p_{x}$, we simulate the transmitted excitation intensities during the first ($n=1$) up to the $n$-th reflections:
\begin{align*}
I_1^{exc}&=I_0^{exc}[p_{0w}+F(1-A)p_{0sw}]R_w\\
I_2^{exc}&=I_1^{exc}[p_w+p_s(1-A)p_{sw}]R_w\\
I_3^{exc}&=I_2^{exc}[p_w+p_s(1-A)p_{sw}]R_w\\
\vdots\\
I_n^{exc}&= I_{n-1}^{exc}[p_w+p_s(1-A)p_{sw}]R_w\\
&=I_0^{exc}[p_{0w}+F(1-A)p_{0sw}]R_w\\
&\times \big\{[p_w+p_s(1-A)p_{sw}]R_w\big\}^{n-2}. 
\end{align*}
Here, $p_{0sw}$ and $p_{sw}$ indicate the probabilities of light passing through the sample and hitting the wall for the first and consecutive reflections, respectively. Similarly, we evaluate the absorbed intensity ($I^{abs}$) by the sample/reference and the intensity recorded by the detector ($I^{det}$) at the exit of the IS by: 
\begin{align*}
I_1^{abs}&=I_0^{exc}F A\\
I_{n}^{abs}&=I_{n-1}^{exc}p_sA \\
I_1^{det}&=I_0^{exc}[p_{0d}+F(1-A)p_{0sd}]\\
I_{n}^{det}&=I_{n-1}^{exc}[p_d+p_s(1-A)p_{sd}].
\end{align*}

For the emitted light intensity ($I^{em}$) and its fraction recorded by the detector ($I^{det^*}$), we consider a different reflectivity of the IS coating $R_w^*$ and an effective sample absorptance $A^*$:
\begin{align*}
I_1^{em}&=I_0^{em}p_{0w}^*R_w^*\\
I_n^{em}&=I_{n-1}^{em}R_w^*[p_w+p_s(1-A^*)p_{sw}]\\
I_1^{det^*}&=I_0^{em}p_{0d}^*\\
I_n^{det^*}&=I_{n-1}^{em}[p_d+p_s(1-A^*)p_{sw}].
\end{align*}

To account for re-absorption and subsequent re-emission, $A^*$ is defined as $A^*=A(\lambda_{em})(1-\eta)$, i.e.\ the fraction that is absorbed by the sample, but not re-emitted. The initial emission intensity originating from the sample is given by $I_0^{em}=I_{tot}^{abs}c\eta$, where $I_{tot}^{abs}$ is the total excitation intensity absorbed in the sample and $c$ is the fraction of light absorbed by the emitters in the sample (i.e. $c<1$ when the emitters are dispersed in an absorbing matrix or solvent). The total absorbed intensity by the sample during the measurement is calculated by summation of $I_{n}^{abs}$ over all reflection-steps. Using the geometric series, $\sum_{n=0}^{\infty} x^{n} = \frac{1}{1-x}$, we obtain: 
\begin{widetext}
\begin{eqnarray}
I_{tot}^{det}&=&\sum_{n=1}^{\infty}I_n^{det}=I_0^{exc}\Big\{p_{0d}+F(1-A)p_{0sd}+R_w[p_{0w}+F(1-A)p_{0sw}]\frac{p_d+p_s(1-A)p_{sd}}{1-R_w[p_w+p_s(1-A)p_{sw}]}\Big\}\label{det1}\\
I_{tot}^{abs}&=&\sum_{n=1}^{\infty}I_n^{abs}=I_0^{exc}A\Big\{F+R_w[p_{0w}+F(1-A)p_{0sw}]\frac{p_s}{1-R_w[p_w+p_s(1-A)p_{sw}]}\Big\}\label{abs1}\\
I_{tot}^{det^*}&=&\sum_{n=1}^{\infty}I_n^{det^*}=I_0^{em}\Big\{p_{0d}^*+p_{0w}^*R_w^*\frac{p_d+p_s(1-A^*)p_{sd}}{1-R_w^*[p_w+p_s(1-A^*)p_{sw}]}\Big\}.\label{det2}
\end{eqnarray}
\end{widetext}
For the spectral sensitivity correction factors $C$ and $C^*$ we assume an empty sphere ($A=0$) and compare the theoretical intensity, equal to the input intensity at the sphere entrance $I_0^{exc}$, with the detected intensity at the sphere exit. This is done separately for the excitation and emission wavelengths, where in the latter case we replace $R_w$ by $R_w^*$.
\begin{align}
C&=\frac{I_0^{exc}}{I_{det}^{tot}(A=0)}=\nonumber\\
&\left[p_{od}+F p_{0sd} + R_w(p_{0w} + F p_{0sw})\frac{p_d+p_s p_{sd}}{1-R_w(p_w+p_s p_{sw})}\right]^{-1}.
\end{align}

Assuming that the reference sample does not emit ($N_{Ref}^*=0$), the QY is given by:
\begin{eqnarray}
QY=\frac{N_{em}}{N_{abs}}=\frac{\int I_{tot}^{det^*}(A_S)C^*\text{d}t}{\int [I_{tot}^{det}(A_{Ref})-I_{tot}^{det}(A_S)]C\text{d}t}
\label{eq:simQY}
\end{eqnarray}

For common IS conditions, such as a non-absorbing reference ($A_{Ref}=0$), no re-absorption ($A_S^*=0$) and inserted baffles ($p_{0d}=p_{0sd}=p_{0d}^*=p_{sd}=0$), the number of emitted $N_{em}$ and absorbed $N_{abs}$ photons in Equation \eqref{eq:simQY} can be expressed as:
\begin{align}
N_{em}=&\int I_0^{exc} \left[F+(p_{0w}+F(1-A_S)p_{0sw}) M p_s\right] A_S \eta \nonumber\\
&\times\frac{p_{0w}^* M^* p_d}{(p_{0w}+ F p_{0sw}) M_{cal}^* p_d}\text{d}t=N_S^*\label{eq:emission}\\ 
N_{abs}=&\int I_0^{exc} \left[1- \frac{\left(p_{0w}+F (1-A_S)p_{0sw} \right) M p_d}{(p_{0w}+ F p_{0sw}) M_{cal}^* p_d} \right] \text{d}t\nonumber\\
=&N_{Ref}-N_S.\label{eq:nabs}
\end{align}
The parameters $M$ and $M^*$ are `sphere-multipliers', defined as $M=R_w(1-R_w[p_w+p_s(1-A)p_{sw})^{-1}$ and $M^*=R_w^*(1-R_w[p_w+p_s(1-A)p_{sw})^{-1}$ and describe how light is distributed over the IS interior and the objects inside it \cite{Valenta2014a}. For the calibration measurements, $M_{cal}=M(A=0)$ and $M_{cal}^*=M^*(A=0)$. Assuming that loss channels are small, $p_{0w}=1-F-p_{0l}\sim 1-F$, Equations \ref{eq:emission} and \ref{eq:nabs} reduce to the QY descriptions found elsewhere \cite{Valenta2014a,Faulkner2012,DeMello1997}. In addition, the validity of our analytical approach has been separately verified using ray-tracing simulations.

\section{Results}\label{results}
\begin{figure}
\includegraphics[width=0.4\textwidth]{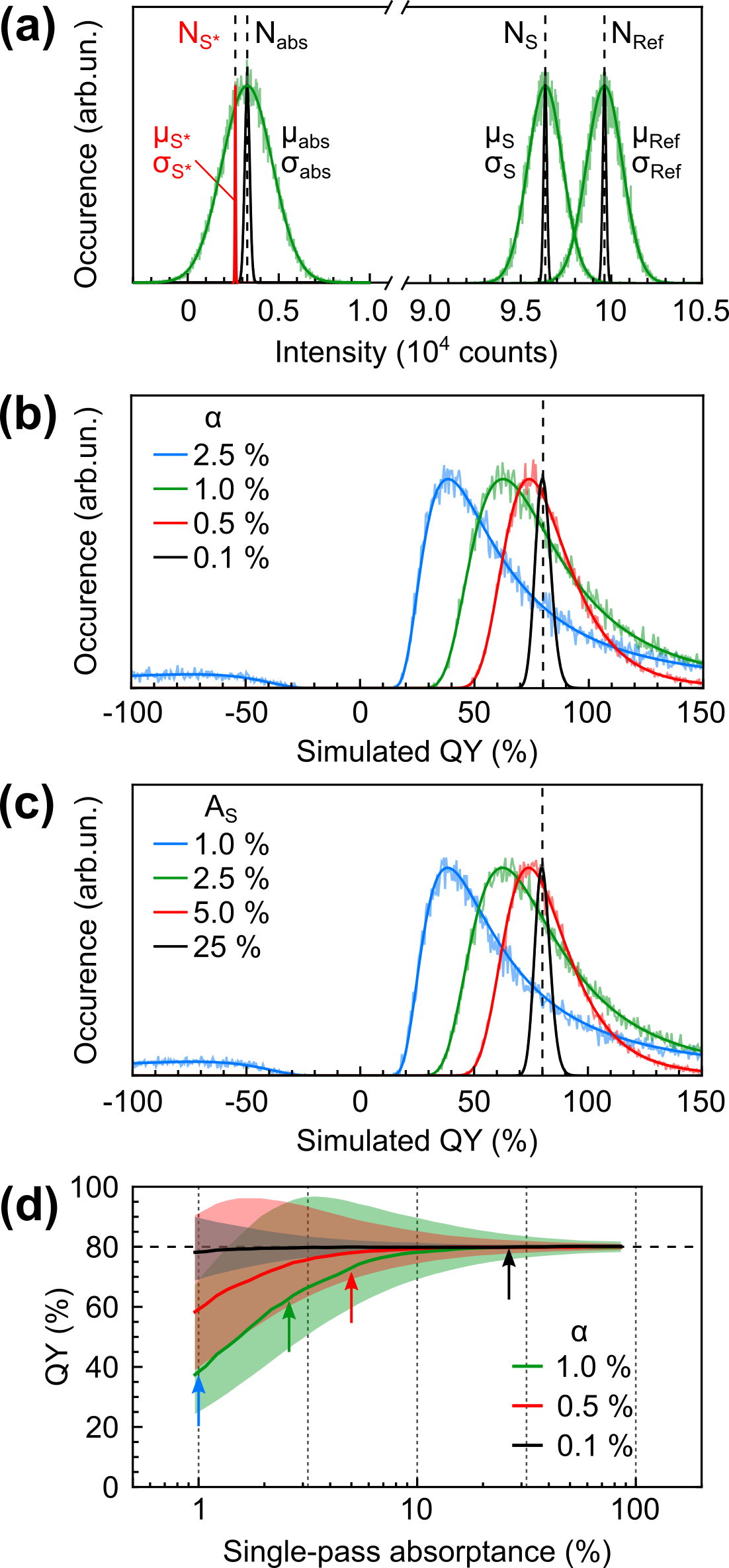}
\caption{(Color online) Simulated effect of normally distributed noise on the QY: (a) Normalized histograms of the number of detected emission photons $N_{em}$ (red) and excitation photons $N_S$ and $N_{Ref}$ for a sample absorptance of 2.5\% (black and green). The latter are shown for a relative measurement uncertainty of 1\% (green) and 0.1\% (black). The thick solid lines show normal distributions; the vertical dashed lines indicate the (noiseless) most-likely values. (b) Normalized histograms of the simulated QY values for a fixed sample absorptance  ($A_S\sim$~2.5\%) and different relative measurement uncertainties and (c) for a fixed measurement uncertainty ($\alpha=$1\%) and a varying sample absorptance. The thick solid lines in (b,c) indicate fits by the analytical expression for the ratio-distribution (Equation~(\ref{prob1})). (d) The simulated most-likely QY value (lines) and the full-width at half-maximum (FWHM) of the positive part of the simulated QY distribution (shaded regions) against the single-pass absorptance of the sample. The horizontal dashed line shows the simulated QY without added noise. Relative uncertainties are set to $\alpha=$~1, 0.5 and 0.1\%. The arrows indicate the absorptance values used in panel (c)} \label{figure2}
\end{figure}

Using the procedure outlined above, we simulate the QY that would be measured in a typical QY geometry, for which we select an IS with a diameter of 10~cm, a reflectance at the excitation and emission wavelengths of 0.97 and 0.99 and set the diameter of the sample, input port and output to 1~cm, 4~mm and 1~mm respectively. The reference sample is assumed to be non-absorbing and non-emitting ($A_{Ref}=0$ and $N_{Ref}^*=0$) and the sample emission efficiency is set to an arbitrary value of $\eta=80\%$. To account for measurement uncertainties in the number of detected excitation and emission photons, we describe those variables by a distribution function with an expectation value determined by Equations (\ref{det1}) and (\ref{det2}) (Figure~\ref{figure2}a). The peak of the distribution indicates the most-likely value of the variable, whereas the standard deviation of the distribution $\sigma$ indicates its fluctuations. 

To cover commonly encountered measurement uncertainties, we discuss two types of distributions: A Poisson and a normal distribution. The Poisson distribution is given by $P(k,\mu)=\frac{\mu^ke^{-\mu}}{k!}$, where $\mu$ is the expected value and $P(k,\mu)$ the probability of measuring a photon count value $k=0,1,2,3,...$ . It is used to describe shot noise, which arises from the discrete nature of photons and will show especially for low flux signals, since the signal-to-noise ratio of the Poisson distribution increases with the square root of the number of detected photons ($\sqrt{N}$). The normal distribution is given by $G(k,\mu,\sigma^2)=\frac{e^{-\frac{(k-\mu )^2}{2 \sigma ^2}}}{\sqrt{2 \pi \sigma^2} }$, where $\mu$ is again the expected (mean) value, $\sigma^2$ is the variance  and $k$ is the variable, i.e. photon counts. The normal distribution can be used to model measurement uncertainties that arise from e.g.\ mechanical/electronic stability of the detection and excitation chains and describes a more general situation in which the variance $\sigma^2$ can be set independently from the expectation value $\mu$. 

In typical QY experiments, the recorded counts $N_{Ref}$ and $N_S$ are usually very high ($10^5$ photon counts or more) and can otherwise easily be increased, by extending the measurement time or by doing multiple runs of the same measurement. As a consequence, the scenario in which (Poissonian) shot noise dominates the signal is unlikely in practice. Moreover, for higher photon fluxes the Poisson distribution can be well approximated by a normal distribution $P(k,\mu)\sim G(k,\mu,\mu)$. We therefore choose to discuss the fluctuations resulting from a normal distribution of the measured variables, to describe a more general source of experimental uncertainty in QY measurements. We will confine the specific case of Poisson distributed variables to the Supplementary Materials.

To study the effect of measurement uncertainty on the QY, we simulate the distribution of the number of detected photons $N_S$, $N_S^*\equiv N_{em}$ and $N_{Ref}$ by drawing semi-randomly from a normal distribution, $N_S\sim G(k, \mu_S,\sigma_S^2)$, $N_S^*\sim G(k,\mu_{S^*},\sigma_{S^*}^2)$ and $N_{Ref}\sim G(k,\mu_{Ref},\sigma_{Ref}^2)$. We set the standard deviation of each variable by choosing a fixed value of the relative uncertainty, $\alpha$, defined as $\alpha=\frac{\sigma}{\mu}$, i.e.\ the fluctuation in the measured variables relative to the mean value. The distributions are illustrated for $\alpha=1$\% (green) and $\alpha=0.1$\% (black) in Figure~\ref{figure2}a. Using Equation~(\ref{eq1}) we then compute the QY distribution, shown for different values of $\alpha$ in Figure~\ref{figure2}b. For a low uncertainty $\alpha$ (black curve), the simulated distribution of the number of absorbed photons $N_{abs}=N_{Ref}-N_S$ is narrow, which leads to a QY distribution that lies symmetrically around the expected QY value (dashed vertical line). Upon increasing $\alpha$ (green curve), however, the distribution of $N_{abs}$ broadens with one tail approaching zero, and the QY distribution ($QY\propto 1/N_{abs}$) becomes asymmetric. In this case, the most-likely QY value (the peak of the distribution) is shifted towards lower, underestimated values.  Moreover, there is a finite probability of finding negative QY values when $N_S > N_{Ref}$. A similar effect occurs when $\alpha$ is fixed and the absorptance is varied, as shown for in Figure~\ref{figure2}c. For an absorptance of 1\%, the most-likely QY estimate is underestimated by more than a factor of $\sim2$, i.e. 200\%. The absorption value for which the QY is underestimated strongly depends on the relative uncertainty of the measurement as shown in Figure~\ref{figure2}d. For $\alpha=0.1\%$, the most-likely QY agrees very well with the expected QY, independent of the absorption of the sample. However, already for the relative uncertainties of 0.5\%, the QY estimate is reliable only when the sample's absorptance exceeds $\sim$5\%. For even higher uncertainties of 1\% , the absorption limit is as high as $\sim$15\%. 

\section{Discussion}

\begin{figure}
\includegraphics[width=0.4\textwidth]{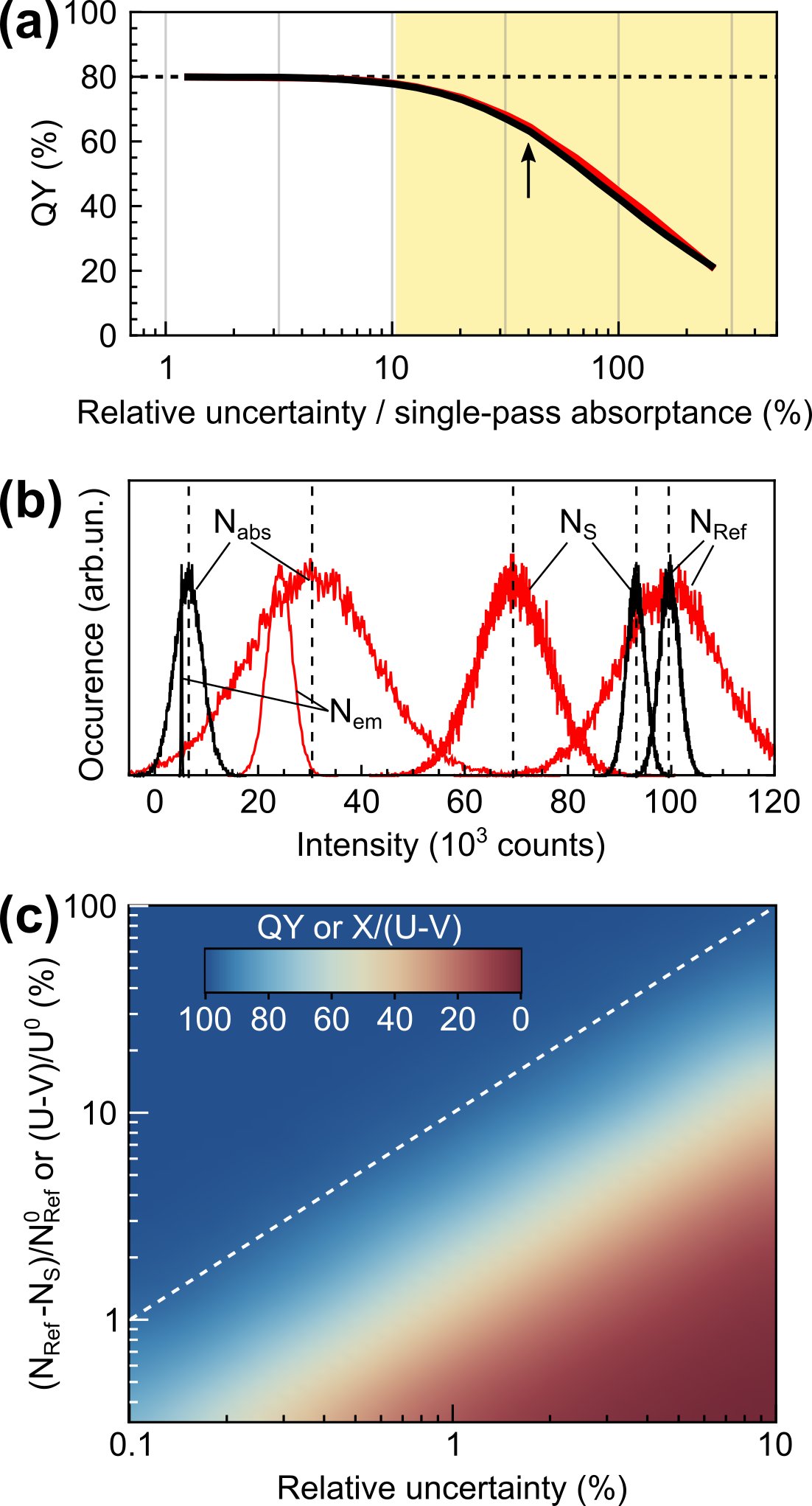}
\caption{(Color online) (a) Simulated most-likely QY value against the relative measurement uncertainty, normalized to the sample's single-pass absorptance $A_S$ for $A_S=$ 5\% (black) and $A_S=$ 25\% (red). The horizontal dashed black line indicates the noiseless QY value. For a relative noise level of $ > 10\%$ of the single-pass absorptance value, the QY is underestimated (yellow area). The arrow indicates the situation in panel (b). (b) Simulated distribution of $N_{Ref}$, $N_S$ and $N_{abs}=N_{Ref}-N_S$ for $A_S=$ 5\% and $\alpha=$ 2\% (black) and $A_S=$ 25\% and $\alpha=$ 10\% (red). (c) QY (color bar) as a function of the normalized number of absorbed photons $(N_{Ref}-N_S)/N_{Ref}^0$ and relative uncertainty in the $N_{Ref}$ and $N_S$ estimates. The dashed white line indicates the threshold below which the QY is unreliable. This holds for any quantity given by the general relation $X/(U-V)$.}\label{figure3} 
\end{figure}

Figure~\ref{figure2}b-d shows that the QY distribution becomes skewed when the fluctuations in the number of detected photons become comparable to the one tenth of the absorptance of the sample: $\alpha /A_S \sim$ 10\%. This results in a large variance in the number of absorbed photons ($N_{abs}$, Figure~\ref{figure2}a), which due to the inverse proportionality $QY \propto 1/N_{abs}$ results in an skewed distribution function of the QY. The relative fluctuations in the number of emitted photons, given by the same $\alpha$, are small (Figure~\ref{figure3}b). Moreover, since $N_{em}$ is in the nominator of the QY definition (Equation~(\ref{eqQYbasic})), it therefore has a negligible effect on the shape of the QY distribution. The full dependence of the QY on the relative uncertainty and absorptance is shown in Figure~\ref{figure3}a for samples with a single-pass absorptance of 5\% (black) and 25\% (red). When $\alpha$ is small compared to the absorptance, $\alpha/A_S < 10\%$, the most-likely QY is in good agreement with the unbiased result, whereas for $\alpha/A_S > 10\%$ (yellow area) the QY is increasingly underestimated. For $\alpha$ equal to the sample's single-pass absorptance, the most-likely QY is already reduced to 50\% of the unbiased value. A nearly identical curve is obtained for an arbitrary absorption value, even when the absorptance is high (red curve). In both cases, there is a large variance in the distribution of $N_{abs}$ (Figure~\ref{figure3}b. This shows that it is the ratio of the relative uncertainty and the sample absorption that determines the underestimation of the QY. 

An another important finding it that in the ideal case, in which noise is absent, the IS methodology itself does not yield biased results. The bias (Figure~\ref{figure2}) arises only when accounting for the measurement uncertainty in the simulated number of detected photons predicted by Equations~(\ref{det1}) and (\ref{det2}). Hence we conclude that the bias results purely from the relation from which the QY and, in particular, the absorption part of the QY is determined: $QY\propto 1/(N_{Ref}-N_S)$. The same effect is therefore expected to show for any quantity with the general form $Z=X/Y$ and $Y=U-V$, when the uncertainty in $U$ and $V$ is comparable to the value of $(U-V)$.

We can derive the analytical expression for the probability distribution of this general case. By taking normally distributed random variables $X\sim G(k,\mu_X,\sigma_X^2), U\sim G(k,\mu_U,\sigma_U^2)$ and $V\sim G(k,\mu_V,\sigma_V^2)$, we get $Y\sim G(k,\mu_Y,\sigma_Y^2)$ with $\mu_Y=\mu_U-\mu_V$ and variance $\sigma_Y^2=\sigma_U^2+\sigma_V^2$. It then follows that the derived ratio $Z=X/Y$ is a continuously distributed random variable with the probability density function (for more details, see the Supplementary Materials)\cite{blejec}:

\begin{eqnarray}
p_Z(z)&=&\frac{\theta}{\pi(z^2+\theta^2)}\nonumber\\
&\times&[\sqrt{2\pi}B(z)\Phi(B(z))e^{-\frac{C(z)}{2}}+K]\label{prob1},
\end{eqnarray}
where 
\begin{align*}
B(z)=\frac{\alpha_Y z+\alpha_X\theta}{\alpha_X\alpha_Y\sqrt{z^2+\theta^2}}
\end{align*}
\begin{align*}
\Phi(z)=\int_{-\infty}^z \frac{1}{\sqrt{2\pi}}e^{-\frac{1}{2}u^2}du
\end{align*}
\begin{align*}
C(z)=\frac{(\alpha_Y\theta-\alpha_X z)^2}{\alpha_X^2\alpha_Y^2(z^2+\theta^2)}
\end{align*}
\begin{align*}
K=exp(-\frac{\alpha_X^2+\alpha_Y^2}{2\alpha_X^2\alpha_Y^2}).
\end{align*}

Here, we again define the relative uncertainty $\alpha_i=\sigma_i/\mu_i$ with $i=X,Y,U,V$ and the parameter $\theta=\sigma_X/\sigma_Y$. For simplicity, we assume that the measurements of variables $X, Y, U$ and $V$ are all independent, which typically holds for QY measurements. The general case with dependent variables is discussed in the Supplementary Materials. 

The first factor in Equation~(\ref{prob1}) is the `standard' part of the density of a non-centered Cauchy distribution, which is independent of $\mu_{X,Y}$ and has no mean or variance. The factor in brackets is known as the `deviant' part, and leads to the skewed shape of the distribution. Using the analytical expression in Equation~(\ref{prob1}), we can very precisely fit our simulated QY distributions in Figure~\ref{figure2}b,c (colored full lines), thus validating our results.\\  

Finally, in Figure~\ref{figure3}c we summarize the limitations of the determination of the QY (or any analogous quantity $Z=X/(U-V)$) by plotting the general dependence of the most-likely QY ($Z$) on the relative measurement uncertainty $\alpha$ and the absorptance $A_S$ (or general analogue to absorptance $(U-V)/U^0$), i.e.\ the fraction of the number of incident photons $N_{Ref}^0$ that is absorbed by the sample, $(N_{Ref}-N_S)/N_{Ref}^0$. The threshold below which the QY estimate becomes unreliable is designated by the white dotted line in Figure~\ref{figure3}c and corresponds to the yellow area in Figure~\ref{figure3}a. As a rule of thumb, the QY can be reliably estimated for a sample with an absorptance that exceeds $\sim$10\% of the relative measurement uncertainty, e.g. for an absorptance $>1\%$ when $\alpha=0.1\%$. The QY becomes unreliable upon decreasing absorption or increasing measurement uncertainty. 

\section{Conclusion}
In conclusion, we report on a general issue arising in experimental methodologies where a quantity in the form $Z=X/Y$ with small $Y=U-V$ is evaluated, such as the photoluminescence QY. These quantities are biased towards lower values when $(U-V)/U^0$ is $\sim 10\%$ of the uncertainty in the variables $U$ and $V$. This happens commonly in QY measurements of low-absorbing samples, since for typical noise levels of $\sim$~1\%, absorptance below 10\% is already affected. Those numbers are not uncommon in many experiments, which suggests that this bias could have been present in published results that rely on the QY methodology \cite{Faulkner2012, Timmerman2011, Miller2012, Sun2015}. This artifact has passed undetected for a long time, due to the assumption that the error in the measured signal intensities propagates in a linear manner and therefore that low photon fluxes, e.g. when measuring low-absorption or emission materials, merely result in a larger uncertainty in the obtained QY value \cite{Valenta2014a, chung2015investigating}. The underestimation arises purely from the uncertainty in the measured variables $U$ and $V$ compared to the normalized difference $(U-V)/U^0$. Hence we anticipate that, not only the absolute QY, but also the comparative and relative QY techniques will suffer from this effect. By detailed numerical simulations and by derivation of the skewed probability density function of the QY we quantify this effect and provide guidelines for the range of absorption values for which the QY can be reliably determined in each specific experimental setup.

\section*{Author contributions}
Analytical model was co-developed by BB, KD and BvD. BvD carried out simulations. GD derived the general analytical expression for the ratio $Z=X/Y$. The manuscript was written, through contributions of all authors. All authors have given approval to the final version of the manuscript. 

\begin{acknowledgments}
BvD, BB and KD acknowledge Dutch STW funding, FOM Projectruimte No. 15PR3230 and MacGillavry Fellowship. GD acknowledges support from the ESIF, EU Operational Programme Research, Development and Education, and from the Center of Advanced Aerospace Technology (CZ.02.1.01/0.0/0.0/16\_019/0000826), Faculty of Mechanical Engineering, Czech Technical University in Prague. Also, the authors thank J. K\v{r}iv\'{a}nek, A. Wilkie and I. Kondapaneni (Faculty of Mathematics and Physics, Charles University) and A. Perez for fruitful discussions. 
\end{acknowledgments}

\section*{Supplementary Materials}

\subsection{Possion distribution}
\begin{figure}
\includegraphics[width=0.4\textwidth]{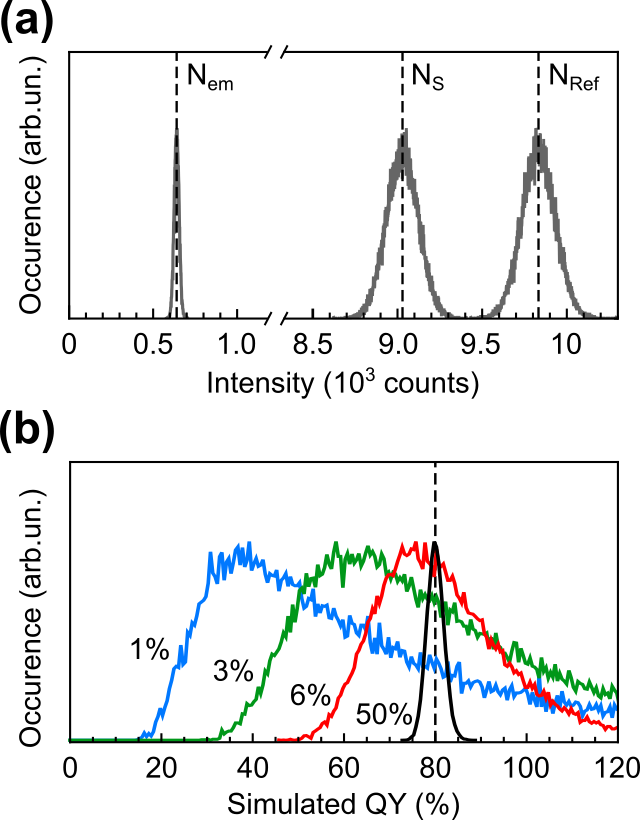}
\caption{(Color online) Simulated effect of Poisson distributed noise on the QY: (a) Normalized histograms of the detected number of emission ($N_{em}\equiv N_S^*$) and excitation photons $N_S$ and $N_{Ref}$ for a sample absorptance of 6\%. The vertical dashed lines indicate the expected, noiseless quantities. (b) Normalized histograms of simulated QY values for different sample absorptance. When the absorptance is high, the distribution of obtained QY values lies symmetrically around the noiseless QY value of 80\% (dashed line). For decreasing absorptance, the distribution shifts towards lower QY values and becomes asymmetric. A total number of $\sim 10^7$ excitation photons was used for the simulations.} \label{figureS1}
\end{figure}

The Poisson distribution is given by $P(k,\,\mu)=\frac{\mu^k}{k!}e^{-\mu}$, where $\mu$ is the expected value and $P(k,\,\mu)$ the probability of measuring a (photon count) value $k=0,1,2,3,...$ . The Poisson distribution is used to describe shot noise, which arises from the discrete nature of photons. This type of noise will show especially for low flux signals since the signal-to-noise ratio increases with $\sqrt{N}$. To simulate the effect of Poisson noise, we add this noise to the simulated numbers of detected photons $N_S$, $N_S^*(\equiv N_{em})$ and $N_{Ref}$, by drawing semi-randomly from a Poisson distribution, $P(k,\,N_S)$, $P(k,\,N_S^*)$ and $P(k,\,N_{Ref})$, as illustrated in Figure~\ref{figureS1}a. We vary the relative uncertainty in the variables $\alpha=\sigma/\mu$, which goes with $\sim 1/\sqrt{N}$ for a Poisson distribution, via the single-pass absorptance of the sample ($A_S$) and the total number of excitation photons ($N_0^{exc}=\int I_0^{exc} dt$).\\

\begin{figure}
\includegraphics[width=0.4\textwidth]{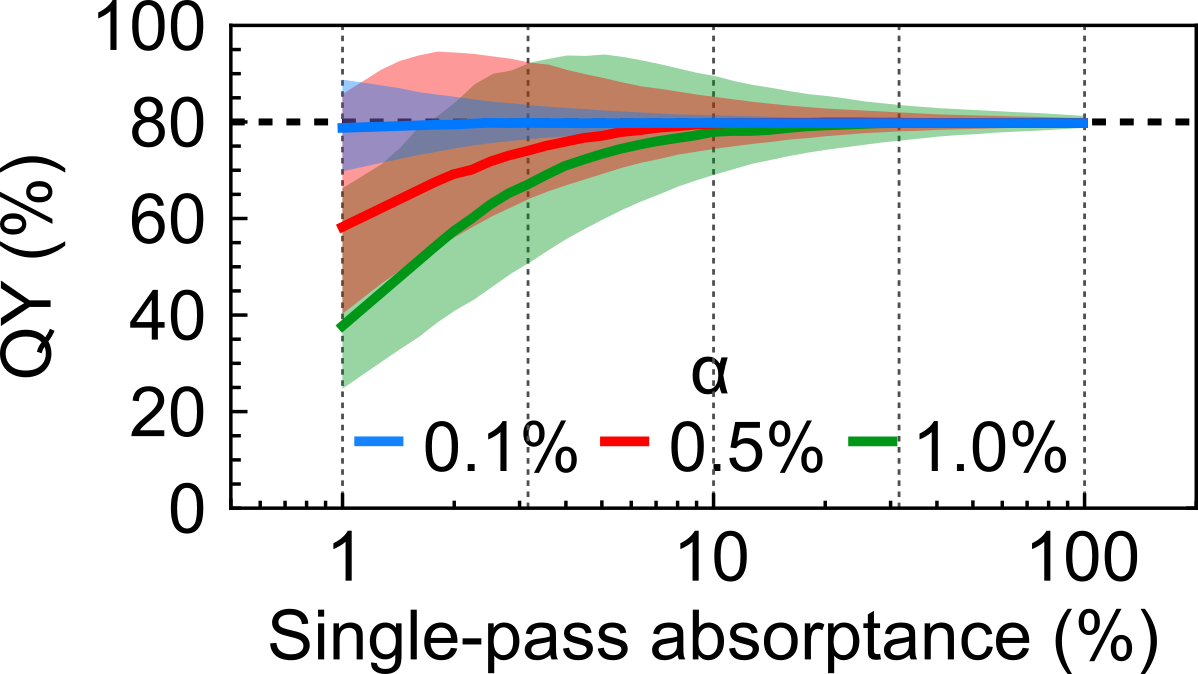}
\caption{(Color online) (a) The simulated most-likely QY (the peak value) and the full-width at half-maximum (FWHM) of the positive part of the simulated QY distribution against the single-pass absorptance of the sample. The relative uncertainty is varied via the number of excitation photons $N_{Ref}^0$: $\sim 10^7$, $5 \cdot10^7$ and $10^9$. The horizontal dashed line shows the simulated QY without added noise.}\label{figureS2}
\end{figure}

The resulting QY distributions are shown in Figure~\ref{figureS1}b and~\ref{figureS2}. For a relative uncertainty of 1\% and for high sample absorptance, the simulated QY distribution lies symmetrically around the expected QY value as shown in Figure~\ref{figureS1} (dashed vertical line). However, upon decreasing absorptance, the distribution broadens and becomes asymmetric. The most-likely QY value (the peak of the distribution) shifts towards lower, underestimated values. For an absorptance of 1\%, the most-likely QY estimate is underestimated by more than a factor of $\sim2$, i.e. 200\%. The absorptance value for which the QY is underestimated strongly depends on the relative uncertainty of the measurement as shown in Figure~\ref{figureS2}. For $\alpha=0.1\%$, obtained under a high number of excitation photons $N_{exc}\equiv N_{Ref}^0 \sim10^9$, the most-likely QY agrees very well with the expected QY, independently of the absorptance of the sample. However, already for relative uncertainties of 0.5\% ($N_{Ref}^0=5\cdot 10^7$), the QY estimate is reliable only for absorptance above $\sim$ 5\%. For even higher uncertainties of 1\% ($N_{Ref}^0\sim10^7$), the absorptance limit is as high as $\sim$~15\%.\\

These results are very similar to those found for normal distributed variables (Figure~2 in the main text). Indeed, for higher fluxes, the Poisson distribution can be well approximated by a normal distribution by setting the variance $\sigma^2$ equal to the expectation value: $G(k,\mu,\sigma^2)=G(k,\mu,\mu)$. Hence, for a fixed value of $\alpha$ any differences in our simulations between Poisson and normal distributions are expected to disappear. For low flux with $\mu\ll 50$, however, the Poisson distribution is clearly asymmetric, from which an additional bias could arise as was shown by \textit{Park et al.}\ \cite{Park2006}. \\

\bigskip
\subsection{Ratio distribution}
Let $X,Y$ be two continuously distributed random variables with a joint probability density function $p_{XY}(x,y)$. Then the ratio $Z=\frac{X}{Y}$ is a continuously distributed random variable with a probability density function \cite{marsaglia}:
\begin{equation}
p_Z(z)= \int_{-\infty}^{\infty}\,|y|\,p_{XY}(zy,y)\,\rm dy\label{eqS1}
\end{equation}
 
\bigskip
For normally distributed random variables $X$,$Y$ with a bivariate normal density function, with a vector mean $(\mu_X, \mu_Y)^T$ and a covariance matrix with marginal standard deviations $\sigma_X^2, \sigma_Y^2$ and a correlation coefficient $\rho_{XY}$, we can derive the particular form of this density function \cite{fieller}:

\begin{align}
p_Z(z)= \frac{b(z)d(z)}{a^3(z)\sigma_X\sigma_Y\sqrt{2\pi}}\left[2\Phi\left\{\frac{b(z)}{a(z)\sqrt{1-\rho^2}}\right\}-1\right] \nonumber \\
+ \frac{\sqrt{1-\rho^2}}{a^2(z)\pi\sigma_X\sigma_Y}
\rm{exp}\left\{-\frac{c}{2(1-\rho^2)}\right\},\label{eqS2}
\end{align}

where 
\begin{eqnarray*}
a(z) &=&\left(\frac{z^2}{\sigma_X}-\frac{2\rho z}{\sigma_X\sigma_Y}+\frac{1}{\sigma_Y^2}\right)^{\frac{1}{2}},\\
b(z) &=& \frac{\mu_Xz}{\sigma_X^2}-\frac{\rho(\mu_X+\mu_Yz)}{\sigma_X\sigma_Y}+\frac{\mu_Y}{\sigma_Y^2},\\
c &=& \frac{\mu_X^2}{\sigma_X^2}-\frac{2\rho\mu_X\mu_Y}{\sigma_X\sigma_Y}+\frac{\mu_Y^2}{\sigma_Y^2},\\
d(z) &=& \rm{exp}\left\{\frac{b^2(z)-ca^2(z)}{2a^2(z)(1-\rho^2)}\right\}.
\end{eqnarray*}
Note that
$$
\Phi(w)=\int_{-\infty}^w \phi(u)\rm d u, \ \ \text{where} \ \ \phi(u)=\frac{1}{\sqrt{2\pi}}\rm{e}^{-\frac{1}{2}u^2}
$$
and the bivariate normal density has the form
\begin{widetext}
\begin{align*}
p_{XY}(x,y)=\frac{1}{2\pi\sigma_X\sigma_Y\sqrt{1-\rho^2}}\rm{exp}\left(-\frac{1}{2(1-\rho^2)}\left[\frac{(x-\mu_X)^2}{\sigma_X^2}+\frac{(y-\mu_Y)^2}{\sigma_Y^2}-\frac{2\rho(x-\mu_X)(y-\mu_Y)}{\sigma_X\sigma_Y}\right]\right\}.
\end{align*}
\end{widetext}
 
The density in Eq.~\ref{eqS2} can be written in the following product form \cite{blejec}:
\begin{align}
p_Z(z)=&\frac{\sigma_X\sigma_Y\sqrt{1-\rho^2}}{\pi(\sigma_Y^2z^2-2\rho\sigma_X\sigma_Yz+\sigma_X^2)}\nonumber \\
&\times\left[\sqrt{2\pi}B(z)\Phi\left(B(z)\right)\rm{exp}\left(-\frac{C(z)}{2}\right)+K\right],\label{eqS3}
\end{align}
where 
\begin{eqnarray*}
B(z) &=& \frac{(\sigma_Y^2\mu_X -\rho\,\sigma_X\sigma_Y\mu_Y)z - \rho\,\sigma_X\sigma_Y\mu_X + \sigma_X^2\mu_Y}{\sigma_X\sigma_Y\sqrt{(1-\rho^2)(\sigma_Y^2z^2-2\rho\,\sigma_X\sigma_Yz+\sigma_X^2)}},\\
C(z) &=& \frac{(\mu_X-\mu_Yz)^2}{\sigma_Y^2z^2 - 2\rho\,\sigma_X\sigma_Yz + \sigma_X^2},\\
K &=& \rm{exp}\left(-\frac{\sigma_Y^2\mu_X^2 -2\rho\,\sigma_X\sigma_Y\mu_X\mu_Y + \sigma_X^2\mu_Y^2}{2\sigma_X^2\sigma_Y^2(1-\rho^2)}\right).
\end{eqnarray*}

\bigskip
Denoting  $\alpha_X=\sigma_X/\mu_X$, $\alpha_Y=\sigma_Y/\mu_Y$ and  $\theta=\sigma_X/\sigma_Y$, we can rewrite Eq.~\ref{eqS3} as
\begin{align}
p_Z(z)=&\frac{\theta\sqrt{1-\rho^2}}{\pi(z^2-2\rho\theta z+\theta^2)}\nonumber\\
&\times\left[\sqrt{2\pi}B(z)\Phi\left(B(z)\right)\rm{exp}\left(-\frac{C(z)}{2}\right)+K\right],\label{eqS4}
\end{align}
where 
\begin{eqnarray*}
B(z) &=& \frac{(\alpha_Y -\rho\alpha_X)z + (\alpha_X - \rho\alpha_Y)\theta}{\alpha_X\alpha_Y\sqrt{(1-\rho^2)(z^2-2\rho\,\theta z+\theta^2)}},\\
C(z) &=& \frac{(\alpha_Y\theta -\alpha_X z)^2}{\alpha_X^2\alpha_Y^2(z^2 - 2\rho\,\theta z + \theta^2)},\\
K &=& \rm{exp}\left(-\frac{\alpha_X^2 -2\rho\alpha_X\alpha_Y + \alpha_Y^2}{2\alpha_X^2\alpha_Y^2(1-\rho^2)}\right).
\end{eqnarray*} 

The first factor in Eq.~\ref{eqS4} is the density of a non-centered Cauchy distribution (or `standard part') $C(\phi, \tau)$ with a location parameter $\phi=\rho\,\theta$ and a parameter of variability $\tau=\theta\sqrt{1-\rho^2}$. The location parameter $\phi$ represents both the modus and the median. The Cauchy distribution has no mean and no variance. Note that the standard part is independent of $\mu_X$ $\mu_Y$.  \\

The second factor in the brackets is the `deviant part', which is a strictly positive and asymptotically constant function $D(z)$ \cite{blejec}: 
$$
\lim_{z\to \pm\infty}D(z)=const(\,\rho, \alpha_X, \alpha_Y, \theta\,)
$$    

In the quantum yield model, we suppose that $X$ is independent of $Y$, which implies $\rho=0$. Hence we obtain the final form of the probability density as:   
\begin{equation}
p_Z(z)=\frac{\theta}{\pi(z^2+\theta^2)}\left[\sqrt{2\pi}B(z)\Phi\left(B(z)\right)\rm{exp}\left(-\frac{C(z)}{2}\right)+K\right],\label{eqS5}
\end{equation}
where 
\begin{eqnarray*}
B(z) &=& \frac{\alpha_Y z + \alpha_X\theta}{\alpha_X\alpha_Y\sqrt{z^2+\theta^2}},\\
C(z) &=& \frac{(\alpha_Y\theta -\alpha_X z)^2}{\alpha_X^2\alpha_Y^2(z^2 + \theta^2)},\\
K &=& \rm{exp}\left(-\frac{\alpha_X^2 + \alpha_Y^2}{2\alpha_X^2\alpha_Y^2}\right).
\end{eqnarray*} 

\begin{figure}[h]
\includegraphics[width=0.45\textwidth]{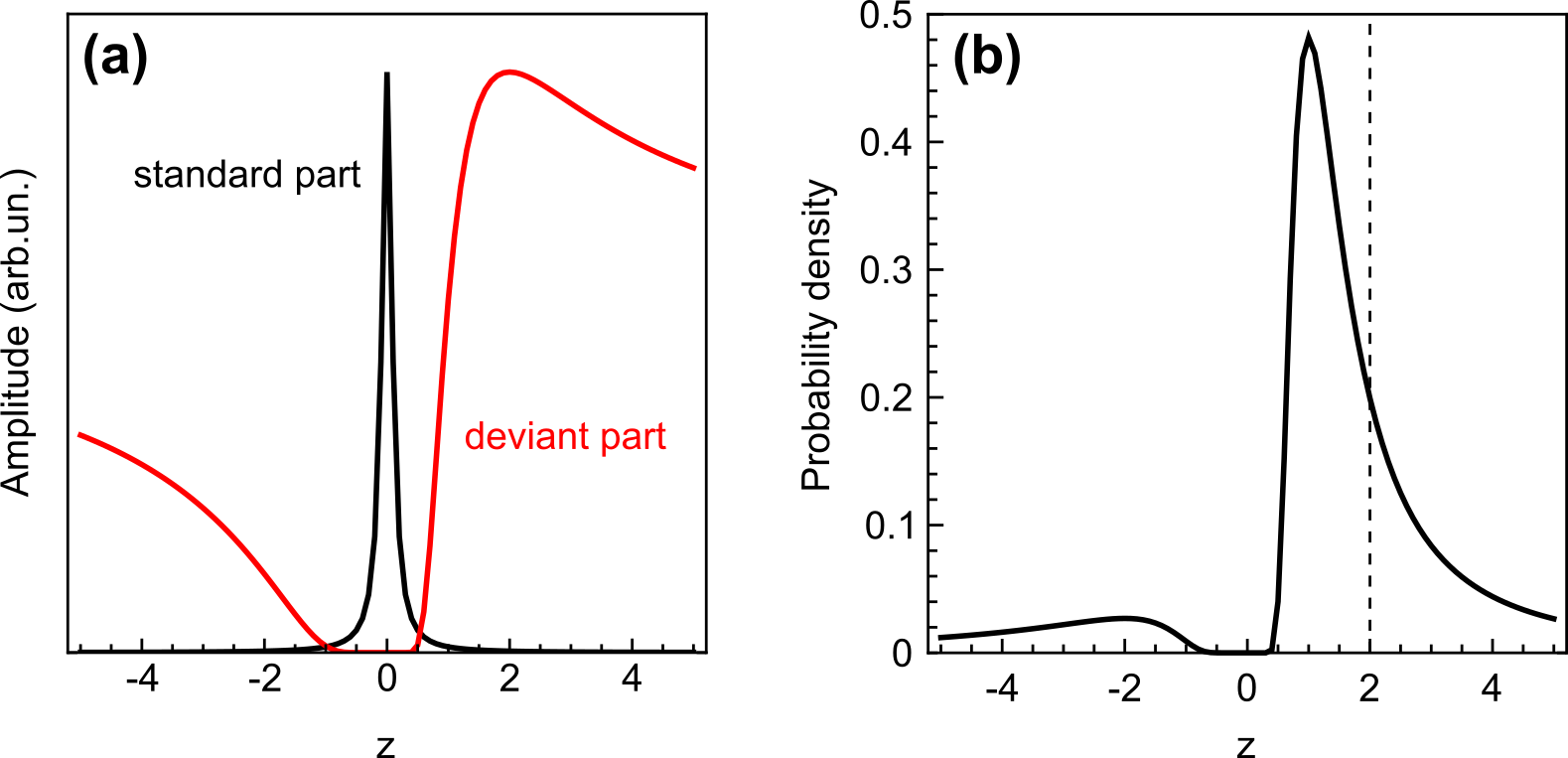}
\caption{An example of the ratio distribution density for $X\sim G(k,2,0.1)$, $Y\sim G(k,1,1)$ and $\rho=0$. (a) The decomposition into the standard part (black), i.e.\ the density of Cauchy distribution, and the deviant part (red). (b) The resulting probability density of the ratio $Z=X/Y$.}\label{figureS3}
\end{figure}

An example of the probability density $p_Z(z)$ and its decomposition into the standard and deviant part is shown in Figure~\ref{figureS3}. 

Since in many case the distribution is bimodal, we can find the maximum by omitting the negative values of $Z$. For this we denote $Z^+= Z$ if $Z\geq 0$ and $Z^+=0$ whenever $Z<0$. Then we should use a truncated distribution $p_{Z^+}(z)=T_r\,p_{Z}(z)$ for $z\geq 0$ and $p_{Z^+}(z)=0$ for $z<0$. Here $K$ is the constant $T_r=(1-F_Z(0))$, where $F_Z(z)$ is the cumulative distribution function of $Z$, i.e. $F_Z(z)=\int_{-\infty}^z p_Z(x)\text{d} x$. Importantly, the shape of the truncated density is similar to the original density. Among other it means that modus (location parameter) remains the same.

\subsection{Experimental test}
\begin{figure*}[ht]
\includegraphics[width=0.6\textwidth]{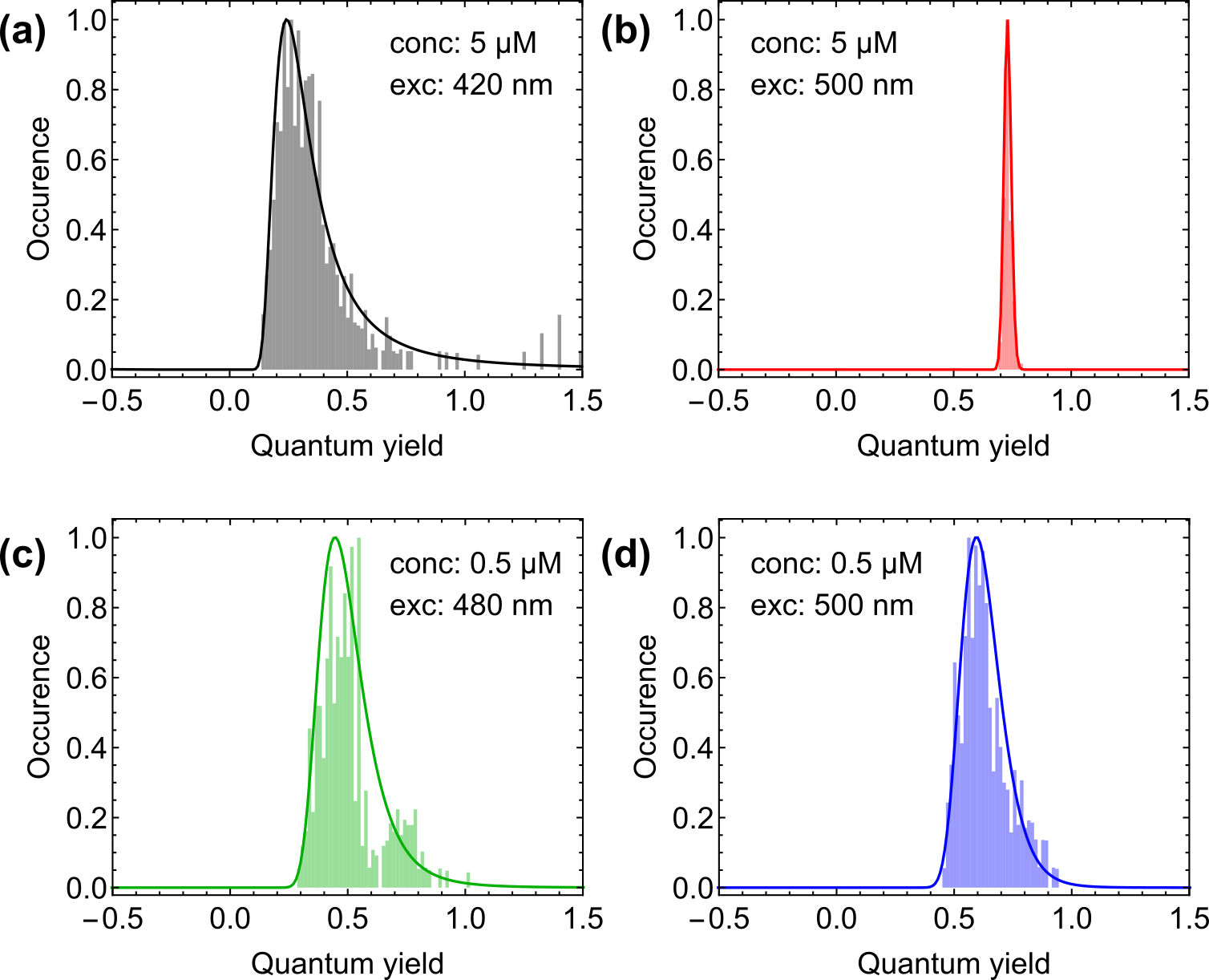}
\caption{Histogram of measured QY values of Rhodamine 6G in ethanol for different values of the single-pass absorptance (black: 0.6\%; red, 20\%; green: 0.8\%, blue: 2\%), varied via the sample concentration and excitation wavelengths (see the legends). The solid lines show the ratio distribution described in Equation~\ref{eqS5}.}\label{figureS4}
\end{figure*}

To test our theoretical model, we performed series of measurements of $N_S$, $N_{Ref}$ and $N_S^*$ using Rhodamine 6G (R6G) in ethanol. To obtain a varying sample absorptance, we prepared two different concentrations, 0.5 and 5$\mu$M and used three different excitation wavelengths of 420 nm, 480 nm and 500 nm. The resulting histograms of the QY values obtained from $N_S$, $N_{Ref}$ and $N_S^*$ via $QY=N_{em}/N_{abs}=N_S^*/(N_{Ref}-N_S)$, are shown in Figure~\ref{figureS4}. The QY distribution for the absorptance of 20\% is shown in Figure~\ref{figureS4}b, obtained for a sample with a concentration of 5$\mu$M under 500 nm wavelength excitation, which is close to the excitation spectrum peak of R6G. In this case, the QY distribution is narrow and centered around $QY\sim$75\%. The absorptance is lowered either by keeping the excitation wavelength and diluting the sample (Figure~\ref{figureS4}d), or keeping the concentration and changing the excitation wavelength away from the excitation band (Figure~\ref{figureS4}a), or both (Figure~\ref{figureS4}c). For a lower sample absoprtance, the QY histograms become broad, skewed, with a most-likely (peak) value further away from the noiseless expected value observed for the higher absorption in panel (b). QY histograms are in good agreement with the probability density predicted by Equation~\ref{eqS5} (or Equation~10 in the main text of the manuscript). 

\section*{REFERENCES}

%

\end{document}